%% file: preprint.tex
\documentclass[aps,preprint, prl,showpacs,superscriptaddress,endfloats]{revtex4-1}  % for review and submission
\usepackage{graphicx}% Include figure files
\usepackage{dcolumn}% Align table columns on decimal point
\usepackage{bm}
\usepackage{multirow,rotating} % for sideways labels in table...
\usepackage{color}
\usepackage{comment}

\def\figAsize{12cm}
\def\figBsize{16cm}
\def\figCsize{16cm}
\def\figDsize{16cm}
\newcommand{\reft}[1]{Table \ref{#1}}

\begin{document}
\input{main-prl4}
\end{document}

%% file: main-prl4.tex
\title{Connections between Human Dynamics and Network Science}

\author{Chaoming Song}
\affiliation{\footnotesize{Center for Complex Network Research,
Department of Physics, Biology and Computer Science, Northeastern
University, Boston, Massachusetts 02115, USA}}
\affiliation{\footnotesize{Center for Cancer Systems Biology, Dana
Farber Cancer Institute, Boston, Massachusetts 02115, USA}}
\author{Dashun Wang}
\affiliation{\footnotesize{Center for Complex Network Research,
Department of Physics, Biology and Computer Science, Northeastern
University, Boston, Massachusetts 02115, USA}}
\affiliation{\footnotesize{Center for Cancer Systems Biology, Dana
Farber Cancer Institute, Boston, Massachusetts 02115, USA}}
\author{Albert-L\'aszl\'o Barab\'asi}
\affiliation{\footnotesize{Center for Complex Network Research,
Department of Physics, Biology and Computer Science, Northeastern
University, Boston, Massachusetts 02115, USA}}
\affiliation{\footnotesize{Center for Cancer Systems Biology, Dana
Farber Cancer Institute, Boston, Massachusetts 02115, USA}}
\affiliation{\footnotesize{Department of Medicine, Brigham and
Women's Hospital, Harvard Medical School, Boston, Massachusetts
02115, USA}}

%\date{\today}

\begin{abstract}
The increasing availability of large-scale data on human
behavior has catalyzed simultaneous advances in network theory,
capturing the scaling properties of the interactions between a
large number of individuals, and human dynamics, quantifying the
temporal characteristics of human activity patterns. These two
areas remain disjoint, each pursuing as
separate lines of inquiry. Here we report a series of generic relationships
between the quantities characterizing these two areas
by demonstrating that the degree and link weight distributions in social
networks can be expressed in terms of the dynamical exponents
characterizing human activity patterns. We test the validity of
these theoretical predictions on datasets capturing various facets
of human interactions, from mobile calls to tweets.
\end{abstract}

%\pacs{89.75.Da, 89.75.Hc, 89.65.Ef}

\maketitle

Fueled by data collected by a wide range of
high-throughput tools and technologies, the study of complex
systems is currently reshaping a number of research fields, from
cell biology to computer science. Nowhere are these advances more
apparent than in the study of human dynamics and social media.
Indeed, the unparalleled use of email, mobile devices and social
networking have provided researchers access to massive amounts of data on the
real time activity patterns of millions of individuals,
simultaneously fueling advances in two research areas, network
science \cite{caldarelli2007book, *cohen2010complex,
*dorogovtsev2003book} and human dynamics
\cite{castellano2009statistical, *rybski2009scaling, *rybski2011communication, *Lazer2009science,
*brockmann2006nature}.
%an active branch of computational social science \cite{Lazer2009science}.
Network science focuses on the structure and dynamics of
complex networks that capture the totality of interactions between
individuals, having led to the discovery of a series of generic
properties of real networks, from the fat tailed nature of
the degree distribution \cite{albert2002statistical,
barabasi1999emergence} to predictable patterns characterizing the
weights or link strengths \cite{goh2001universal,
*barrat2004architecture, *barrat2004weighted}. Human dynamics in
contrast focuses on the temporal aspects of individual interaction
patterns, offering evidence that the interevent time between
consecutive events initiated by an individual follow a fat tailed
distribution \cite{barabasi2005origin, *oliveira2005human,
*vazquez2005exact, *vazquez2006modeling, *gabrielli2007invasion, *gabrielli2009invasion,
*malmgren2009universality, *moro2011socialNetwork,
castellano2009statistical, *rybski2009scaling, *rybski2011communication}, representing a significant deviation from
a Poisson process predicted by random communications. As network
theory \cite{albert2002statistical, %*newman2010networks,
caldarelli2007book, dorogovtsev2003book} and human dynamics
\cite{barabasi2005origin, castellano2009statistical} have
developed in parallel, being pursued as separate lines of inquiry,
we lack relationships between the quantities explored by them,
despite the fact that they often study the same systems and
datasets. In this Letter, we derive a series of scaling relationships that link the quantities characterizing social networks and human dynamics,
and demonstrate their generality across a wide range of systems.
%showing that the widely studied properties uncovered independently in the two areas represent two facets of the same underlying phenomena.

To demonstrate the practical relevance of our results, we compiled
four independent datasets that together capture most aspects of
digital communication that humans are involved in lately (SM
Section 1): 1) \emph{Mobile phone data}, that summarizes the
communication patterns of about 4 million anonymized European
mobile users during a year period, providing access to over $1.2$
billion events, representing information on who talks with whom
and the timing of each call \cite{onnela2007structure}; 2)
\emph{E-mail traffic} within a university, that collects over two
million email messages sent during an 83 day period exchanged by
around 3,000 users \cite{eckmann2004entropy, barabasi2005origin}; 3)
\emph{Twitter data}, that records the tweets of about 0.7 million
users, containing over 8 million messages collected between Aug
2009 and Mar 2010 \cite{cha2010measuring,
*mislove2011understanding, *golder2011diurnal}. 4) \emph{Online
Messages}, that records more than 500,000 messages sent by
approximately 30,000 active users of a Swedish dating site over
492 days \cite{Holme2004socialNetworks, rybski2009scaling, rybski2011communication}.
%[[Many reports on how many hours people spend on twitter, but few about phone or emails.
%Also, the usage highly depends on the demographics -- teens rely heavily on social media,
%while elders not as much.]]
%\note{Americans spent an estimated 2 hours and 12 minutes tweeting and reading tweets on Twitter.com in November 2010}
%http://www.experian.com/blogs/marketing-forward/2011/01/25/americans-spend-2-hours-13-minutes-per-month-on-twitter-com/

%4) \emph{Surface mail} based
%communication records of three famous individuals, A. Einstein, C.
%Darwin and S. Freud, captures pre-electronic human communication
%patterns, but lacks social network information
%\cite{oliveira2005human}.

\begin{figure}
\newpage
\centering \resizebox{\figAsize}{!}{\includegraphics{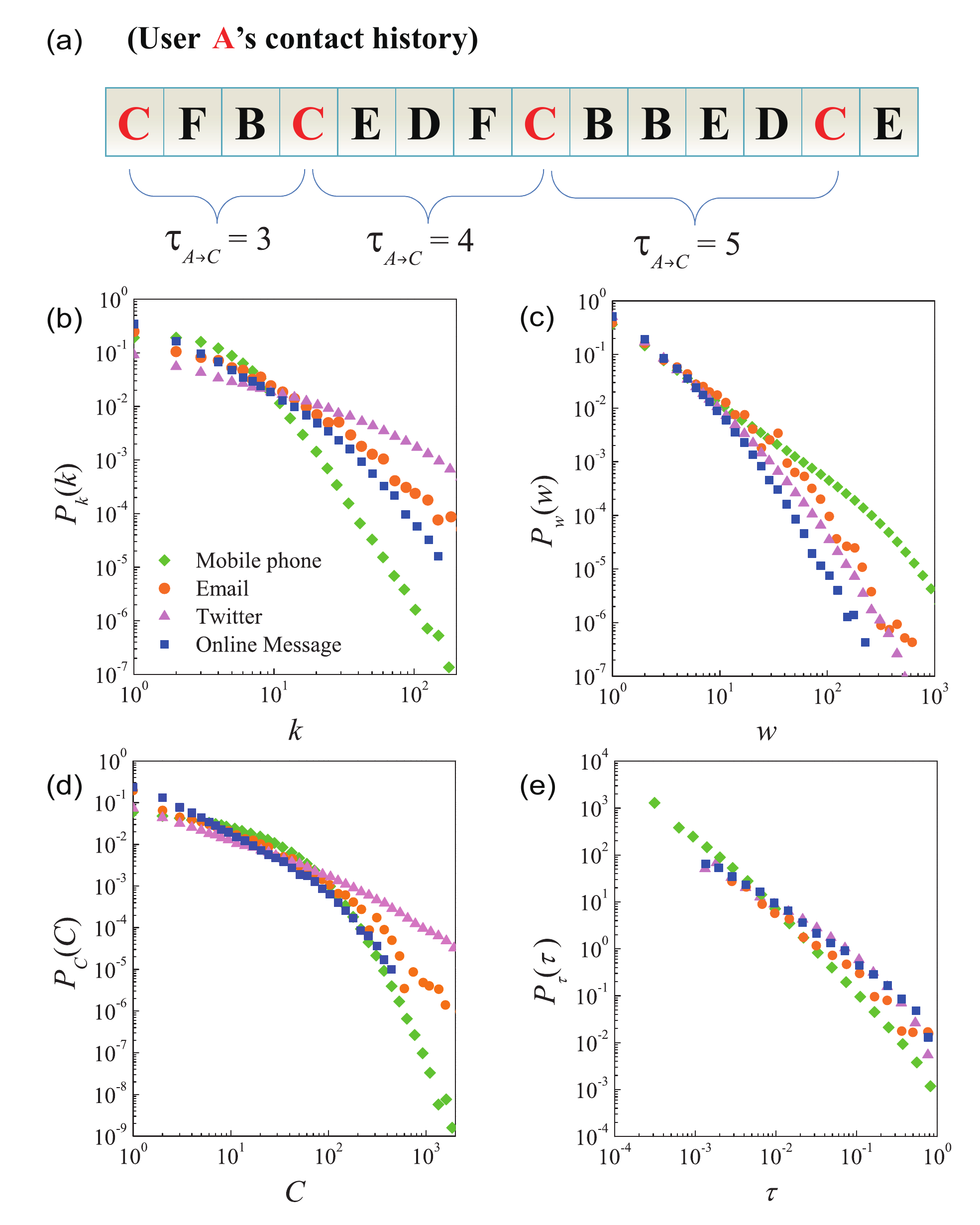}}
\caption{ \textbf{Basic measures characterizing networks and human
dynamics.} (a) The definition of $\tau_{A\rightarrow C}$, the
interevent time captures communication intervals between two
individuals, A and C. Note that $\tau_{A\rightarrow C}$ measures
time in terms of the number of events, a feature that corrects for
daily fluctuations in the communication volume, but has the same
asymptotic scaling as the real interevent time
\cite{eckmann2004entropy}. (b) Degree distribution $P_k(k)$,  and
(c) link weight distribution $P_w(w)$ for each of the four studied
datasets. (d) Activity distribution $P_C(C)$.
(e) The distribution of the number events between
consecutive communications with the same individual,
$P_\tau(\tau)$, where $\tau$ is normalized by each individual's
activity level $C$. \vspace{-2\baselineskip} }\label{fDis}
\end{figure}

Two widely studied quantities characterize the underlying social networks:

\textit{Degree distribution}: The degree $k_i(t_1,t_2)$ of an individual $i$ represents the total number of individuals he/she contacted within the $[t_1, t_2]$ time interval, including both acquaintancy and strong ties \cite{dunbar1992neocortex,*gonccalves2011modeling}. The degree
distribution $P_k(k) \equiv N^{-1}\sum_{i=1}^N \delta(k-k_i)$ of each studied systems can be approximated with a power law \cite{onnela2007structure,
barabasi1999emergence, albert2002statistical} (Fig.~\ref{fDis}b),
\begin{equation}\label{ePk}
P_{k}(k) \sim k^{-\gamma_k},
\end{equation}
where the degree exponent varies between $\gamma_k=1.0$ for
Twitter and $\gamma_k=4.8$ for mobile phones (\reft{tExp}).
The measurements indicate that for Twitter, email, and online
messages $\gamma_k$ is independent of time, but for mobile phones
decreases from $\gamma_k =4.19$ to $\gamma_k =3.20$ during a year
($P_k$ for different time intervals is shown in SM Section 7).

\textit{Weight distribution:} Denoting with $w_{i\rightarrow j}$ (weight) the
number of contacts between two nodes
\cite{goh2001universal, barrat2004architecture,
barrat2004weighted}, we measure the weight distribution $P_w(w) \sim \sum_{i,j} \delta(w-w_{i\rightarrow j})$  for different dataset
(Fig.~\ref{fDis}c), finding that it can be
approximated with (Fig. \ref{fDis}c)
\begin{equation}
P_{w}(w) \sim w^{-\gamma_w},
\end{equation}
where the weight exponent varies between $\gamma_w=1.51$ for
mobile phones and $\gamma_w=1.9$ for emails (\reft{tExp}).

To explore the dynamics of human activity we focus on two
frequently measured quantities \cite{barabasi2005origin, castellano2009statistical, moro2009prl}:

\textit{Activity distribution:} Denoting with $C_i(t_1,t_2)$ the activity, representing the total
number of communications initiated by individual $i$ within a
$[t_1,t_2]$ time interval, we find that the activity distribution $P_C(C) \equiv N^{-1}\sum_{i=1}^N \delta(C-C_i)$ is fat tailed, following (Fig.~\ref{fDis}d)
\begin{equation}\label{ePC}
P_{C}(C) \sim C^{-(1+\beta_C)},
\end{equation}
where $\beta_C$ ranges between $0.1$ (Twitter) to $3.38$ (mobile
phones) (Fig. \ref{fDis}d and \reft{tExp}, $P_C$ for different time intervals is shown in SM Section 7).

\textit{Interevent time distribution:} A key property of human
dynamics is the non-Poissonian nature of the interevent time
$\Delta t$ between consecutive communication patterns
\cite{barabasi2005origin,oliveira2005human,cobham1954priority,
vazquez2005exact, vazquez2006modeling, gabrielli2007invasion, gabrielli2009invasion}.
Previous studies have found that $P_{\Delta t}(\Delta
t) \sim \Delta t^{-\beta_0}$, with
$\beta_0 \simeq 1$ (SM Section 3.1 and
Refs.~\cite{barabasi2005origin,oliveira2005human,cobham1954priority,
vazquez2005exact, vazquez2006modeling, gabrielli2007invasion,
gabrielli2009invasion}). As $P_{\Delta t}(\Delta t)$ characterizes the
communications between \textit{all} friends, here we define a
link-specific interevent time $\tau_{i \rightarrow j}$ as the
total number of communication events initiated by user $i$ between
two consecutive communications from $i$ to $j$ \cite{karsai2012universal}. %(see SM for more detailed definition).
For example, $\tau_{A \rightarrow C} = 3, 4, 5$ in Fig.~\ref{fDis}a.
We measure the probability density function $P_{\tau, i}(\tau)$ across all individuals, finding they that all follow broad distributions (see SM Section 6).
In Fig.~\ref{fDis}e, we plot $P_{\tau}(\tau) \equiv N^{-1}\sum_i C_i^{-1}P_{\tau,i}(\tau/C_i)$ (see Fig. S6 for $P_{\tau,i}$ for different $C_i$ activity groups), finding that it is also
fat-tailed, well approximated by (Fig.~\ref{fDis}e)
\begin{equation}
P_{\tau}(\tau) \sim \tau^{-(1+\beta_\tau)},
\end{equation}
where $\beta_\tau$ characterizes the inhomogeneity of the
communication pattern for a pair of users, varying between $0.2$
(online messages) and $0.53$ (mobile phones) (\reft{tExp}).  Queuing models predict, however,
$\beta_\tau = \beta_0 = 0$ (fixed queue length) or $0.5$ (variable
queue length) \cite{barabasi2005origin, oliveira2005human,
cobham1954priority, vazquez2005exact, vazquez2006modeling,
gabrielli2007invasion, gabrielli2009invasion}.

In summary, the underlying social network is
characterized by $P_k(k)$ and $P_w(w)$, while the
communication dynamics by $P_\tau(\tau)$ and $P_C(C)$,
each with its system dependent form. These two classes of
phenomena, and the associated distributions, are treated independently
in the literature \cite{albert2002statistical, %newman2010networks,
caldarelli2007book, dorogovtsev2003book,
castellano2009statistical, Lazer2009science, brockmann2006nature, barabasi1999emergence, goh2001universal,
barrat2004architecture, barrat2004weighted, barabasi2005origin, oliveira2005human,
vazquez2005exact, vazquez2006modeling, gabrielli2007invasion,gabrielli2009invasion,
rybski2009scaling, rybski2011communication, castellano2009statistical}.

%Next we derive a set of relationships between the two phenomena and help uncover a universal quantity that appears to be independent of the means of communication used by an individual.

\begin{figure}
\centering \resizebox{\figBsize}{!}{\includegraphics{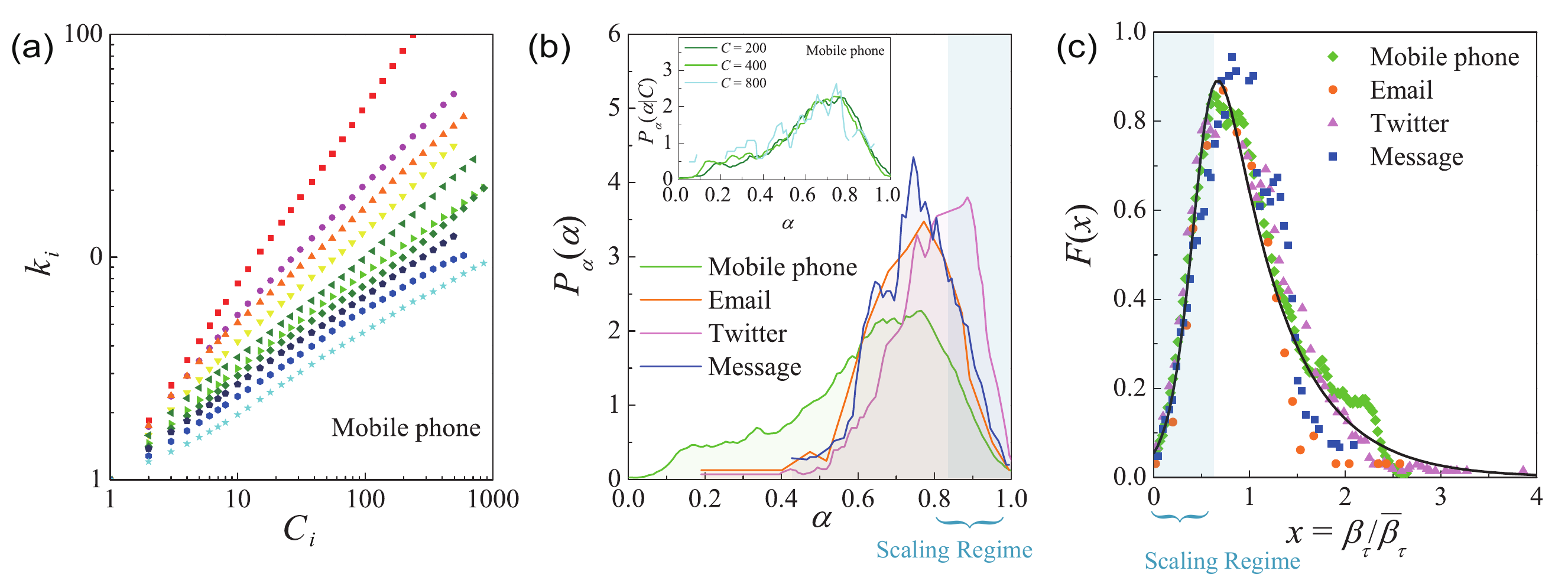}}
\caption{ \textbf{Measuring user sociability.} (a) The growth in
degree $k_i(t_1,t_2)$ for ten mobile phone users in function of the same user's activity $C_i(t_1,t_2)$, where each dot corresponds
$(C_i, k_i)$ for one time frame $[t_1,t_2]$.
Similar curves are observed for the other datasets (see Fig.
S3). (b) The sociability distribution, $P_\alpha(\alpha)$, for the three
studied datasets, where the shaded region highlights the tail of
$P_{\alpha}(\alpha)$.  Inset: conditional probability
distribution $P_\alpha(\alpha|C)$ for mobile phone users with activity $C = 200$, $300$ and $800$,
respectively.
(c) The collapse of
$P_{\beta_\tau}(\beta_\tau)$ distributions after rescaling
$P_{\beta_\tau}(\beta_\tau)$ with average $\overline{\beta_\tau}$
for each datasets.
%The black curve represents a fitting with a Burr
%type II distribution \cite{burr1942cumulative},
%$F(x) \propto \exp(\sigma x)\left(1+ s \exp(\sigma x)\right)^{-(1+\kappa)}$,
%where $\sigma = 0.68 $, $\kappa = 0.24$ and $s = 0.035$.
%Inset: the semi-log plot for small $\beta_\tau$ (shaded area in
%C), capturing the tail of $P(\alpha)$ (shaded area in B).
The black line represents a Burr type II distribution, $F(x)
\propto \exp(\sigma x)/\left(1+ s \exp(\kappa x)\right)$ with
$\sigma = 6.6$, capturing the exponential growth $F(x) \sim
\exp(6.6x)$ for small $\beta_\tau$. \vspace{-1\baselineskip} }
\label{fPKC}
\end{figure}

While one expects that the more active is an individual (high
$C_i$), the more friends he/she has (high $k_i$), as shown in
Fig.~\ref{fDis}b,e and \reft{tExp}, the distributions $P_k(k)$
and $P_{C}(C)$ are not equivalent. To understand the relationship
between $k_i$ and $C_i$, we measured for each individual how their
degree ($k_i$) grows with the number of communication events
($C_i$) they participated in.
%As shown in, where
%we plot the relationship between $C_i$ and $k_i$ for ten randomly
%selected individuals from the mobile phone data, they show widely different growth patterns.
We find that the individual degree $k_i$
can be approximated with (Fig.~\ref{fPKC}a)
\begin{equation}\label{eScaling}
k_i(t_1,t_2) \sim C_i(t_1,t_2)^{\alpha_i},
\end{equation}
where the exponent $\alpha_i$, which characterizes the individual's affinity
to translate its level of activity into new contacts, varies from individual to individual.
For each user $\alpha_i< 1$, the degree grows sub-linearly with the activity $C_i$, indicating diminishing impact on the growth in the number of friends when increasing the number of calls. This is also known as Heaps' law \cite{heaps1978information}, a rather robust phenomenon observed in a broad range of applications and models \cite{barabasi2005origin, oliveira2005human,
cobham1954priority, gnedin2007notes, *krings2012effects}.
While the temporal patterns of both $k_i$ and $C_i$ might
be affected by environmental factors and circadian rhythms, we find that Eq. (\ref{eScaling})
is independent of the observational time frame.

The fact that the exponent $\alpha_i$ varies from individual to
individual indicates that users with similar activity levels
acquire degrees at different rates (Fig.~\ref{fPKC}a). Therefore,
$\alpha_i$ characterizes an individual's ability to add friends
given his/her activity level $C_i$, prompting us to call
$\alpha_i$ \emph{sociability}. To investigate the demographic
variation of sociability, in Fig.~\ref{fPKC}b we show the
sociability distribution for all four
datasets, finding that $P_{\alpha}(\alpha) \equiv N^{-1} \sum_i \delta(\alpha-\alpha_i)$ is bounded between 0 and 1
and decays rapidly on both sides of the peak.
%The variation in
%sociability across the population suggest a multifractal nature of
%human society, where $P(\alpha)$ serves as the multifractal
%spectrum \cite{stanley1988multifractal}.
We also find that $\alpha_i$ is largely independent of $C_i$, as
indicated by the conditional probability $P_{\alpha}(\alpha|C)$,
that overlaps for users with different activity $C$
(Fig.~\ref{fPKC}b, inset). Somewhat surprisingly, this indicates
that sociability, i.e.~the ability to establish new contacts, is largely
independent of the individual's activity level, representing
instead an intrinsic property of an individual. Figure \ref{fPKC}b shows $P_{\alpha}(\alpha)$ for all datasets,
indicating that each communication system is characterized by its
own distinct $P_{\alpha}(\alpha)$ and average sociability
$\overline{\alpha}$ (see \reft{tExp}).

The sociability $\alpha_i$ is related to the dynamical exponent $\beta_{\tau,i}$ as well.
Intuitively, a large $\beta_{\tau,i}$ implies abundance of repeated communications with old contacts,
i.e.~smaller interevent time, corresponding to a slower growth (smaller $\alpha_i$) of an individual degree in the social network.
Indeed, it is easy to show that these two exponents obey (see SM Section 6.1),
%Heaps' law (\ref{eScaling}) allows us to link a node's degree, a network measure, to the bursty patterns of human activity.
%Indeed, let us denote with $\Pi_{i}(t_1,t_2)$ the
%probability that individual $i$ contacts a \emph{new} friend $j$,
%i.e. someone that $i$ did not contact in the previous $[t_1, t_2]$ time
%frame. This requires that the waiting time $\tau_{i\rightarrow j}$, that characterizes the communication between $i$ and $j$, be greater
%than $C_{i}(t_1,t_2)$. The probability that the interevent time exceeds $C_{i}(t_1,t_2)$ is $\Pi_{i}(t_1,t_2) = \int_{C_i}^\infty
%P_{\tau,i}(\tau)d\tau \sim C_i^{-\beta_{\tau,i}}$. Yet, (\ref{eScaling}) indicates that $\Pi_i = dk_i/dC_i \sim C_i^{\alpha_i - 1}$.
%Comparing these two equations. Indeed, we obtain the scaling relationship (see SM Section 6.1),
\begin{equation}\label{eAB2}
\alpha_i + \beta_{\tau, i} = 1.
\end{equation}
%Equation (\ref{eAB2}) indicates that the bursty nature of human activity patterns ($\beta_{\tau, i}$) determines
%the growth of individual degree in the social network via $k_i \sim C_i^{1-\beta_{\tau, i}}$, predicting the observed sublinear growth. Therefore, Eq.~(\ref{eAB2}) represents a link between human dynamics (e.g.~$\tau_{i\rightarrow j}$ and $\beta_{\tau, i}$) and the structure of the underlying social network (e.g. $k_i$ and $\alpha_i$).
%Equation (\ref{eAB2}) offers a rather strong prediction, indicating that a simple scaling relationship holds for each individual,
%despite the individual variability in $\alpha_i$ and $\beta_{\tau, i}$.
%For validation we need to determine $\alpha_i$ and $\beta_{\tau, i}$ for each individual,
%which requires sufficient individual statistics.
%However, as sociability $\alpha_i$ is largely independent of individual activity level (inset of Fig.~\ref{fPKC}b),
%we can focus on active users, for which we have sufficient data, without introducing a selection bias.
As shown in \reft{tExp} (for sake of simplicity, the average $\overline \alpha$  and $\overline {\beta_\tau}$ are reported) and SM Section 6.1, the prediction (\ref{eAB2}) is not only validated by the exponents measured in each dataset, but also consistent with existing models \cite{barabasi2005origin, oliveira2005human, cobham1954priority, gnedin2007notes, *krings2012effects}. %,vazquez2005exact, vazquez2006modeling, gabrielli2007invasion, gabrielli2009invasion
%We report in \reft{tExp} the averaged $\overline \alpha$  and $\overline \beta_\tau$ for each dataset.
%Queuing models predict
%$\beta_\tau = \beta_0 = 0$ (fixed queue length) or $0.5$ (variable
%queue length)
%We find, however, that the empirically observed $\beta_{\tau,i}$ depends both on particular individual and dataset (see SM XXX).
%Yet, the generic relationship (\ref{ePtau}) are verified at each individual level (see SM XXX). To further test the scaling identity (\ref{eAB2}),
%The fact that $P_{\alpha}(\alpha)$ is independent of activity level then allows us to calculate $P_{\beta_\tau}(\beta_\tau)$ distribution from
%$P_\alpha(1-\beta_\tau)$ (Fig.~\ref{fPKC}c).
Perhaps most surprisingly, we find that when rescaled with the average $\overline{\beta_{\tau}}$, $P_{\beta_\tau}(\beta_{\tau})$
%(calculated from $P_\alpha(\alpha)$)
for the different datasets collapse into a single curve (Fig.~\ref{fPKC}c)
\begin{equation}\label{eCollapse}
P_{\beta_\tau}(\beta_\tau) =
(1/\overline{\beta_{\tau}})F(\beta_{\tau} /
\overline{\beta_{\tau}}),
\end{equation}
suggesting that the distribution $P_{\beta_\tau}(\beta_\tau)$ of the bursty exponent $\beta_{\tau}$ captures an inherent
property of the population, independent of the means of
communication. This data collapse is quite remarkable, given the
difference in the nature of the data (calls, emails, tweets, and
online messages), timeframes, countries and demographics (phone:
about 25\% of an European country's population
\cite{onnela2007structure}; emails: university employees from a
different European country \cite{eckmann2004entropy,
barabasi2005origin}; Twitter: mainly US~\cite{cha2010measuring,
mislove2011understanding, golder2011diurnal}; Online Messages: Swedish teenagers
\cite{Holme2004socialNetworks, rybski2009scaling, rybski2011communication}).
Figure~\ref{fPKC}c suggests an exponential growth of $F(x)$ for
small $x$, i.e., $F(x) \sim \exp(\sigma x)$, where $\sigma \approx
6.6$ appears to be the same for all datasets (\reft{tExp}), a
parameter that will play an important role below.
%The average exponent $\overline {\beta_\tau}$, depending on different communication system, can be extracted independently from
%the average waiting time distribution $P_\tau(\tau)$ (Fig.~\ref{fDis}d), which represent
%a power law with finite cutoff rooted from the population heterogeneity.

\begin{figure}
\centering \resizebox{\figCsize}{!}{\includegraphics{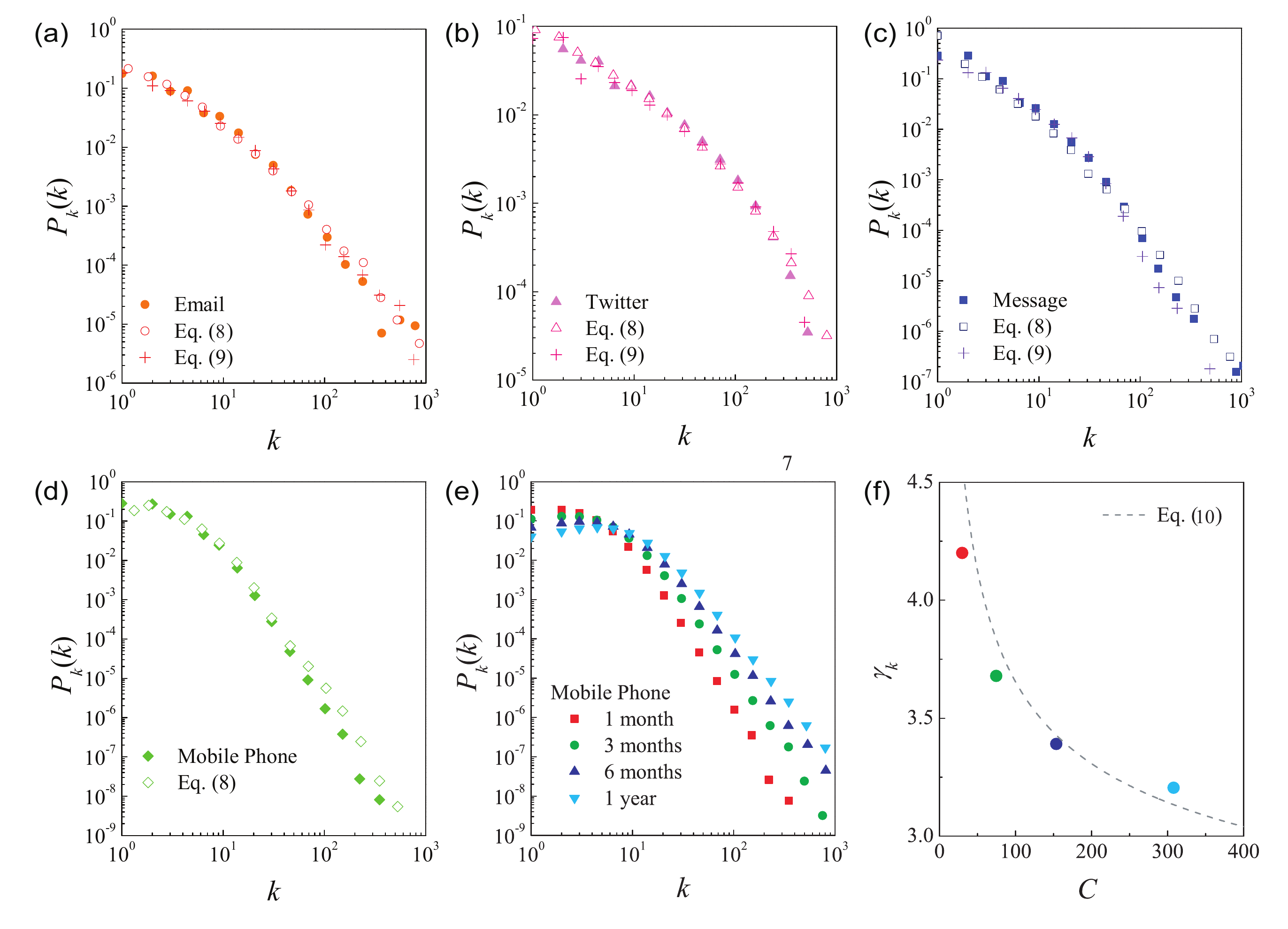}}
\caption{ \textbf{Predicting the Degree Distribution}
The measured degree distribution $P_k(k)$ (solid), compared to the predictions of Eqs. (\ref{ePKsC}) (open) and (\ref{eClass1}) (cross)
for (a) Email (b) Twitter (c) Online Message and (d) Mobile Phone datasets, respectively, showing that Eq. (\ref{ePKsC}) is consistent with empirical observation. For emails we adjusted (\ref{ePKsC}) to allow for multiple recipients (see SM Section 2).
The validation of Eq. (\ref{eClass1}) for Email, Twitter and Online Message datasets also indicates these systems belong to \emph{Case 1}.
%The mobile phone communication, shows a notable deviation from the prediction of Eq. (\ref{eClass1}).
(e) $P_k(k)$ for mobile phone dataset, revealing
power law tails for different time frames $\Delta T \equiv t_2 - t_1$, from 1 month to 1 year (see SM Section 7 for all datasets). (f) The degree exponent $\gamma_k$ deceases with average activity $\overline C$ as predicted by (\ref{eClass2}), indicating that mobile phone communication belongs to \emph{Case 2}.
\vspace{-1\baselineskip} }
\label{fPKsC}
\end{figure}

The scaling law (\ref{eScaling}), together with the sociability distribution $P_{\alpha}(\alpha)$ allows us to
derive an another relationship between social networks and human dynamics.
Indeed, the statistical independence between $\alpha$ and $C$ implies
\begin{equation} \label{ePKsC}
P_{k}(k) = \int \delta(k - C^\alpha) P_{\alpha}(\alpha)P_{C}(C) d\alpha dC,
\end{equation}
indicating that the fat tailed nature of the degree
distribution is rooted in the population heterogeneity in terms of
sociability $\alpha_i$ and activity $C_i$.  Note that this relationship is independent of the particular form of $P_k(k)$ and $P_C(C)$,
being equally valid if they follow power laws, stretched exponentials or log-normal distributions.
We compared the empirically measured $P_k$ with the prediction (\ref{ePKsC}) for all datasets, obtaining excellent agreement (Fig.~\ref{fPKsC}a-d).
Therefore, Eq.~(\ref{ePKsC}) links quantities describing human dynamics ($P_C(C)$) and the social networks ($P_k(k)$), capturing the competition between two phenomena:

\emph{Case 1}: If $P_{k}(k)$ is dominated by differences in the
users' activity level (the activity distribution $P_{C}(C)$),
we can ignore the variations in $P_\alpha$, replacing
individual sociability ($\alpha_i$) with $\overline \alpha$, finding
\begin{equation}\label{eClass1}
P_k(k) \sim k^{1/\overline \alpha-1}P_C(k^{1/\overline \alpha}).
\end{equation}
This limit correctly describes email, twitter, and online messages
(Fig. \ref{fPKsC}a-c).

%$P(\alpha)$ is narrow that one could
%assume all individuals' sociability would be characterized by the
%mean $\overline \alpha$, and accordingly $P_k(k)$ is dominated by
%the activity distribution $P_C(C)$.
%is also independent of the length of the observation time.

\emph{Case 2}: If $P_{\alpha}(\alpha)$ dominates, the individuals'
activity level ($C_i$) can be approximated with their mean
$\overline C$, and Eq.~(\ref{eCollapse}) predicts that the
sociability distribution has an exponential tail $P_{\alpha}(\alpha) \sim \exp(-\alpha \sigma/\bar \beta_\tau))$ (shaded area
in Fig.~\ref{fPKC}b) that dominates the scaling of
(\ref{ePKsC}), obtaining
\begin{equation}\label{eClass2}
P_k(k) \sim k^{-\left(1 +\sigma/(\overline
\beta_\tau\ln \overline C)\right)}.
\end{equation}
This indicates that $P_k$ has a power law tail, whose exponent $\gamma_k$ is determined by variability in sociability,
captured by the parameter $\sigma$. More interestingly, it predicts that $\gamma_k$ decreases with the average activity
level $\overline C$, leading to a scaling
exponent that depends on an extensive quantity, not observed
before in network science. Indeed, as $\overline C$ increases with
the observation time (Fig.~S2), (\ref{eClass2}) predicts a
time-dependent $\gamma_k$, driven by changes in $\overline C$. Figure \ref{fPKsC}e-f show that despite the temporal stationarity of individual activity (Fig.~ S8a) for  mobile communications, $\gamma_k$ decreases with $\overline C$ for different time interval $[t_1, t_2]$, indicating that the degree heterogeneity of mobile phone users is indeed driven by variability in their sociability.

%The distinction between the two classes relies on the variations in individual activity.
%Indeed, Internet-based tools provide opportunity for high-frequency communication, allowing some users to send a significant number of emails/tweets/messeges within short time frames, being activity driven.
%In contrast, mobile phone communication is largely limited by the duration of a call and therefore it is driven by sociability.
%To quantify different social system, we need a model of the activity distribution $P_C(C)$.
%We find that if $P_C(C)$ is bounded (exponential or gaussian), the corresponding systems are always sociability driven;
%if, however, $P_C(C)$ follows a broad distribution, the system reveals a more complicated behavior.
Combining (\ref{ePC}) with these two classes, we predict the degree exponent in (\ref{ePk}), as (SM Section 4)
\begin{equation}
\gamma_k = 1 + \mathrm{min}\left[\frac{\beta_C}{1-\overline
\beta_\tau}, \frac{\sigma}{\overline \beta_\tau\ln
\overline{C}}\right]. \label{eGammak}
\end{equation}
In \reft{tExp} we report the $\gamma_C$ and $\gamma_k$ of the power law model for all datasets.
Yet, Eqs. (\ref{ePKsC}-\ref{eClass2}) are not limited to power laws; other fat tailed models for $P_C$ such as lognormal or stretched exponential can also be exploited, as discussed in SM Section 4. The fundamental relationship (\ref{ePKsC}) and the distinction between the two classes is therefore independent of particular models (and fits) for $P_C$.

\begin{figure}
\centering \resizebox{\figDsize}{!}{\includegraphics{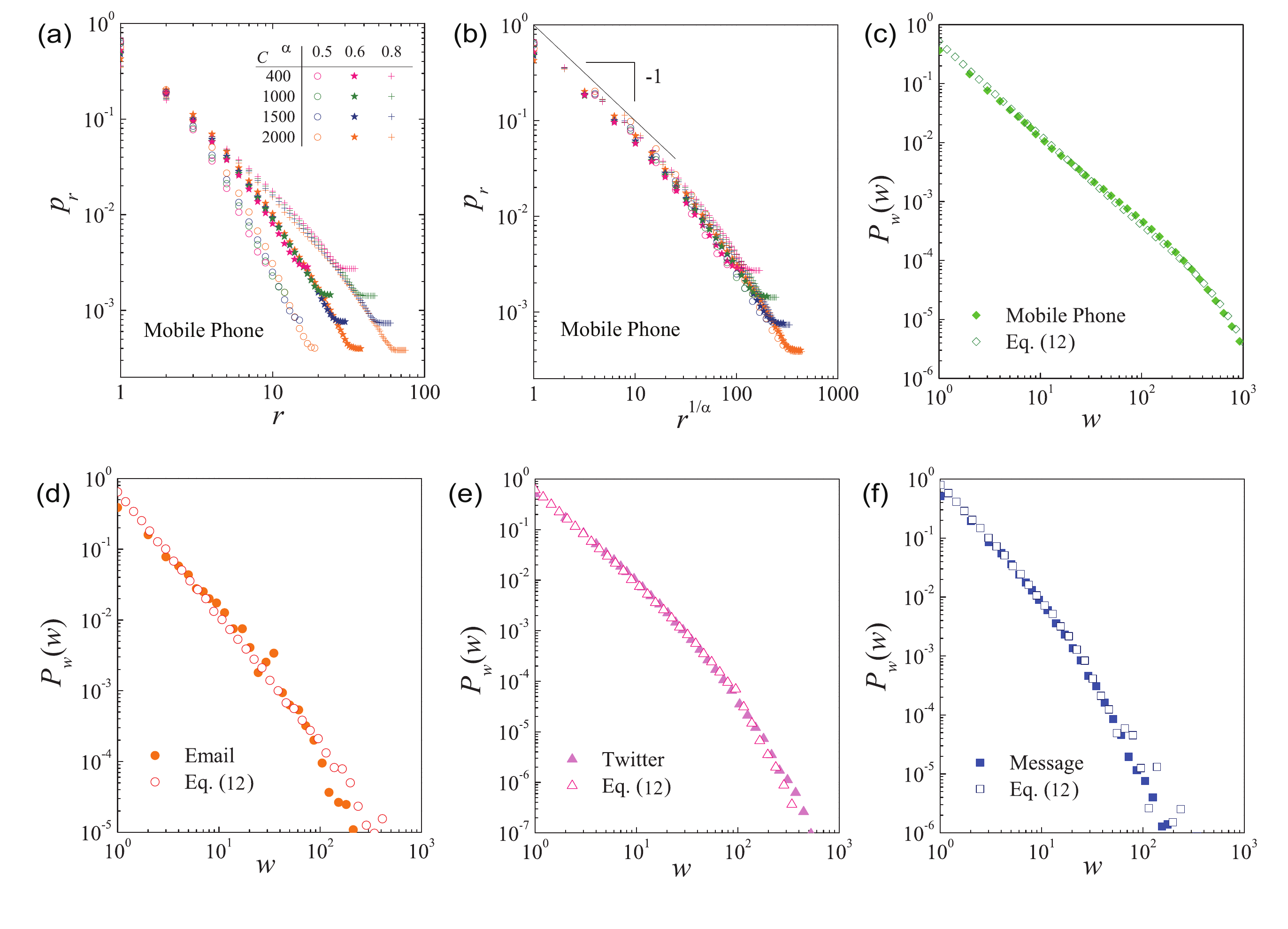}}
\caption{\textbf{Quantifying the tie strength distribution.} (a)
Zipf's plot showing the communication frequency $p_{r,i}$ for a
user $i$ with the user's $r$-th most contacted friend for the
mobile phone data (see the same plot for other datasets in Fig. S4).
The different colors and symbols represent
different activities and sociabilities, respectively, indicating
that the Zipf's exponent $\zeta_i$ depends only on the sociability
$\alpha_i$. (b) The plot of $p_r$ versus $r^{1/\alpha}$ showing
collapses over different sociability groups, as predicted by
$\alpha_i \zeta_i  = 1$. Similar plots are
observed for the other datasets (see SM Section 3.3).
(c,d,e,f) The degree distribution $P_w(w)$ from empirically
measurements (solid), comparing to the predictions of Eq. (\ref{ePw}) for (c) Mobile Phone (d) Email
(e) Twitter and (f) Online Message datasets, respectively, showing that Eq. (\ref{ePw}) is
consistent with the empirical observation.
\vspace{-1\baselineskip} }\label{fZipf}
\end{figure}

To derive the network's weight distribution $P_w(w)$ we
note that for each individual $i$, $\sum_j w_{i\rightarrow j} =
C_i$, where \textit{$w_{i\rightarrow j}$} denotes the total number
of messages/calls from $i$ to $j$. We denote with $p_{r}\equiv p_{i\rightarrow j}
\equiv w_{i\rightarrow j}/C_{i}$ the probability that user $i$
communicates with user $j$, and $r$ is the rank of $p_{i\rightarrow j}$
across all friends $j$ of user $i$. We find that $p_r$ is well approximated by Zipf's law $p_r
\sim r^{-\zeta_i}$  (Fig.~\ref{fZipf}a) \cite{sole2011zipf}, a direct
consequence of the fat tailed nature of $P_w(w)$
\cite{goh2001universal, barrat2004architecture,
barrat2004weighted}. That is, an individual communicates most of
the time with only a few individuals and it interacts with the
rest of its contacts with diminished frequencies. Intuitively, one
would assume $\zeta$ is the same for individuals with the same
activity $C$. Yet, we find that for three randomly selected users,
each with the same activity $C_i = 400$, $p_r$ has different
$\zeta_i$ exponents (Fig.~\ref{fZipf}a). However, for users
with different activities but the same sociability $\alpha$, the
curves are indistinguishable (Fig.~\ref{fZipf}a), hinting the
existence of a link between $\zeta_i$ and $\alpha_i$. This
relationship can be derived by focusing on an individual's least
preferred contact. Intuitively, there are only a few
communications ($O(1)$) between the individual \emph{i} and
his/her least preferred contact, independent of the activity level
$C_i$. Therefore, given $k_i$, the total number of contacts of
individual $i$ is $C_i p_{k_i} = C_i k_i^{-\zeta_i} =
C_i^{1-\alpha_i \zeta_i} = O(1)$, obtaining $\alpha_i \zeta_i  = 1$, in agreement with the previous studies \cite{baeza2000block,
*lu2010zipf}. Here we corroborate this relationship by showing that $p_r
(r^{1/\alpha})$ collapses for all studied datasets for users with different $\alpha_i$ and the
curve has the slope $-1$ for the top ranked contacts (Fig.~\ref{fZipf}b and Fig.~S4).
The scaling identity $\alpha_i \zeta_i  = 1$ allows us to derive the weight distribution
$P_w(w)$. The weight distribution $P_w(w)$ is averaged over populations, as
\begin{equation} \label{ePw}
P_w(w) = \int \sum_{r = 1}^{C^\alpha} \delta(w - A(C, \alpha) C r^{-1/\alpha}) P_C(C)P_\alpha(\alpha) d C d \alpha,
\end{equation}
where the normalization factor $A(C, \alpha) \equiv \mathcal{A} \sum_{r = 1}^{C^\alpha} r^{-1/\alpha}$ with
a system-dependent constant $\mathcal{A}$ corresponding to the average weight.
Figure \ref{fZipf}c-f confirms the validity of Eq. (\ref{ePw}) for all datasets.
The fact that Zipf's law is equivalent with $P_{w,i}(w) \sim w^{-(1+1/\zeta_i)} = w^{-(1+\alpha_i)}$, where $P_{w,i}(w)$ represent the weight distribution of individual $i$, leads to a first order approximation of Eq. (\ref{ePw}) as $P_w(w) \sim w^{-\gamma_w}$, where the exponent $\gamma_w = 1+\overline{\alpha}$ up to the
leading order. Combining
this with (\ref{eAB2}), we find
\begin{equation}
\gamma_w = 2-\overline \beta_\tau. \label{eGammaw}
\end{equation}
The prediction (\ref{eGammaw}) is supported by the empirical data in
\reft{tExp}.

\begin{table*}\centering
%\scriptsize
\caption{\textbf{Quantify networks and human dynamics.}
The scaling exponents characterizing the networks and human
dynamics in the four studied datasets, as well as the most studied human dynamics models.
%Note that the parenthesis indicates the last digit of the confidence level, e.g. , $2.2(7)$ corresponds to $2.27\pm0.01$.
The reported $\overline \alpha$ and $\overline {\beta_\tau}$ represent average values over the population
for empirical data, where $\overline {\beta_\tau}$ is measured from $P_\tau(\tau) \sim \tau^{-(1+\overline {\beta_\tau})}$ as a first order approximation.
The error of $\overline {\beta_\tau}$ and $\beta_C$ are derived from the error of $1 + \overline {\beta_\tau}$ and $1 + \beta_C$, respectively. Note that the small error bars of exponents are due to the large population size. See SI Section 6 for justification of the goodness of fit.
}\label{tExp}
\begin{ruledtabular}
\begin{tabular}{ccccccc}
                  & Mobile phone & Email & Twitter & Message & \multicolumn{2}{c}{Queueing Models}\\
\cline{6-7} &  &  &  & & { \scriptsize{Fixed Length \cite{cobham1954priority}}} & { \scriptsize{Variable Length \cite{barabasi2005origin}}} \\
                  \hline
$\gamma_k$ & $4.19_{\pm0.01} \div 3.205_{\pm0.007}$ &$2.27_{\pm0.01}$ &$1.241_{\pm0.001}$  & $1.624_{\pm0.003}$ & -- & --\\
$\gamma_w$ & $1.51335_{\pm0.00006}$ & $1.637_{\pm0.003}$ &$1.8483_{\pm0.0006}$   & $1.930_{\pm0.002}$ & -- & --\\
\hline
$\overline{\beta_{\tau}}$ & $0.53823_{\pm0.00001}$ &$0.431_{\pm0.002}$ &$0.3162_{\pm0.0001}$  & $0.360_{\pm0.002}$ & $0$ & $0.5$\\
$\beta_C$ & $3.39_{\pm0.01}$ &$0.82_{\pm0.01}$ &$0.147_{\pm 0.001}$   & $0.430_{\pm0.002}$ & -- & --\\
\hline
$\overline \alpha$& $0.58_{\pm0.01}$ &$0.68_{\pm0.02}$ &$0.78_{\pm0.01}$  & $0.70_{\pm0.01}$ & $1.0$ & $0.5$\\
$\sigma$& $6.6_{\pm0.1}$ &$6.8_{\pm0.2}$ &$6.6_{\pm0.1}$   & $6.6_{\pm0.1}$ & -- & --\\
\hline
$\ln \overline {C}$& $3.4 \div 5.9$ &$4.8$ &$5.4$   & $3.0$ & -- & --\\
\end{tabular}
\end{ruledtabular}
\end{table*}

In summary, Eqs. (\ref{ePKsC}) --
(\ref{eGammaw}) offer direct links between human dynamics and
the architecture of social networks, showing that the degree distribution ($P_k$)
and the tie strength distribution ($P_w$) can be expressed in terms of
the dynamical exponents characterizing the temporal patterns in
human activity, like burstiness ($P_{\tau,i}$) and the activity
level ($P_C$).
%The relationship between these two classes of
%exponents is mediated by the parameter $\sigma$, which we find to
%be independent of the communication technology, hence capturing
%an inherent property of human activity.
These relationships bring an unexpected order to the zoo of exponents
reported in \reft{tExp}, showing that they represent different facets of a deeper underlying reality. While a better understanding of the origin of these exponents requires mechanistic models, tailored to the specific communication phenomena, the relationships
(\ref{ePKsC}) -- (\ref{eGammaw}) derived here are independent of the system's details or the
specific communication mechanism, thus all future models that aim
to account for human dynamics and social networks
in a specific system must obey them.
%Hence, the theory developed here represents the skeleton of a joint mathematical theory of human dynamics and social networks.
As our understanding
of human dynamics deepens with the emergence of new and
increasingly detailed data on both human activity patterns and
social networks, such fundamental relationships are expected to
have an increasing value, helping us anchor future models and
offer a springboard towards a deeper mechanistic understanding of
big data, the often noisy, incomplete, but massive datasets that trail human
behavior.

The authors wish to thank A.~Mislove for providing the Twitter
dataset and P. Holme for providing the Online Messages dataset.
This work is supported by the NSF (IIS-0513650); ONR (N000141010968); DTRA
(WMD BRBAA07-J-2-0035 and BRBAA08-Per4-C-2-0033); DARPA (11645021) and the Network Science
Collaborative Technology Alliance sponsored by ARL
(W911NF-09-2-0053).

%
%\vspace{2\baselineskip}
%\noindent {\color{red}\textbf{==========OLD VERSION for Eq. 7 through Eq. 12 ==========}}
%\newpage
\bibliography{act}

%\newpage